\documentclass[runningheads]{llncs}

\usepackage{amsmath,amsbsy,amssymb,amsfonts}
\usepackage{multirow,multicol,booktabs,wrapfig}
\usepackage{graphicx,float,stfloats,subcaption}
\usepackage{pgfplots}
\usepackage{algorithm,algorithmic}
\usepackage{tikz}
\usepackage{makecell}
\usetikzlibrary{shapes.geometric, positioning,matrix, positioning,calc,decorations}
\usepgfplotslibrary{fillbetween, groupplots}

\usepackage[numbers]{natbib}

\newcommand{\bA}{\boldsymbol{A}}

\newcommand{\best}[1]{\textcolor{black}{#1}}

\begin{document}

\title{Detect \& Score: Privacy-Preserving Misbehavior Detection and Contribution Evaluation in Federated Learning}
\titlerunning{Detect \& Score: Privacy-Preserving MD and CE in FL}

\author{Marvin Xhemrishi\thanks{TUM School of Computation, Information and Technology, Technical University of Munich, Germany}, Alexandre Graell i Amat\thanks{Department of Electrical Engineering, Chalmers University of Technology, Gothenburg, Sweden}, and Balázs Pejó\thanks{Department of Networked Systems and Services, BME \& HUN-REN-BME Information Systems Research Group, Budapest, Hungary}}
\authorrunning{M. Xhemrishi, A. Graell i Amat, and B. Pejó}

\institute{}

\maketitle

\begin{abstract}
    Federated learning with secure aggregation enables private and collaborative learning from decentralized data without leaking sensitive client information. However, secure aggregation also complicates the detection of malicious client behavior and the evaluation of individual client contributions to the learning. To address these challenges, QI (Pejo {et al.}) and FedGT (Xhemrishi {et al.}) were proposed for contribution evaluation (CE) and misbehavior detection (MD), respectively. QI, however, lacks adequate MD accuracy due to its reliance on the random selection of clients in each training round, while FedGT lacks the CE ability. In this work, we combine the strengths of QI and FedGT to achieve both robust MD and accurate CE. Our experiments demonstrate superior performance compared to using either method independently. 
\end{abstract}

\section{Introduction}
\label{sec:intro}

In the realm of machine learning, the correlation between the abundance of training data and model accuracy is widely acknowledged. However, the practicality of this relationship is challenged by the fragmented nature of data across diverse entities. To address this, Federated learning (FL) has emerged as a promising paradigm for privacy-preserving decentralized learning.
Unlike centralized learning, which requires aggregating data from all participants into a central server, FL enables multiple clients to train a shared model locally on their private data, exchanging only model updates rather than raw data. However, despite this decentralized design, research has exposed  privacy vulnerabilities\textemdash revealing that sensitive information about the underlying datasets can still be inferred from the shared model updates. 

To mitigate such risks, several privacy-preserving techniques have been proposed, most notably  differential privacy (DP)~\cite{desfontaines2020sok} and secure aggregation (SA)~\cite{mcmahan2016communication}. DP offers formal privacy guarantees, but this often comes at the cost of reduced model utility. In contrast, SA obscures individual updates without degrading model performance, making it an attractive solution for many applications. In essence, SA hides the individual model updates by cryptographically aggregating them, ensuring that only the final aggregated model is visible to the server.

While SA effectively protects privacy by concealing individual client updates, it also introduces a significant limitation: the server can no longer inspect individual contributions. This makes it  considerably more difficult to detect whether a client has performed a poisoning~\cite{tolpegin2020data} or backdoor\cite{bagdasaryan2020backdoor} attack, or to evaluate the relative importance of clients with respect to one other and  the learning  task~\cite{huang2020exploratory}. As a result, traditional misbehavior detection (MD) methods (e.g.,  KRUM~\cite{blanchard2017machine}), contribution evaluation (CE) techniques (e.g., LOO~\cite{wang2020principled}), and joint approaches (e.g., FedSV~\cite{otmani2024fedsv}) cannot be applied under secure aggregation, as they rely on access to individual model updates. Likewise advanced CE methods, such as  Zeno~\cite{xie2019zeno} and Shapley-value based approaches~\cite{rozemberczki2022shapley}, are not compatible with SA. 

To address this challenge, Xhemrishi \emph{et al.} proposed FedGT~\cite{xhemrishi2023fedgt}, a scheme for client MD under FL with SA tailored to the cross-silo setting. In parallel, and Pejo \emph{et al.} introduced QI~\cite{pejo2023quality}, a CE framework designed for  the  cross-device setting under SA. 

\subsubsection*{Contribution} In this work, we extend the schemes in~\cite{xhemrishi2023fedgt}  and~\cite{pejo2023quality}  and integrate their core ideas to develop a novel approach that supports both MD and CE in a cross-silo FL setting under SA.
The proposed scheme leverages spatial and temporal information to enhance both tasks, and empirical results show that it outperforms FedGT, QI, and other relevant baselines in terms of both MD and CE performance. In more details, we upgrade both schemes such that they can leverage more information and, as such, enhance their performance. Particularly, we adapt FedGT to use temporal information (inter-round information) and transform QI such that spatial information can be leveraged (intra-round information).We observe empirically that the transformed QI becomes a quite powerful tools, both for MD and CE.

 \subsubsection*{Organization}

 In Section~\ref{sec:bg} we summarize the corresponding research efforts as well as the original QI and FedGT schemes. 
 In Section~\ref{sec:model} we envision several settings how FedGT and QI can be utilized for MD and CE. 
 In Section~\ref{sec:exp} we detail the corresponding experiments, while 
 in Section~\ref{sec:conc} we conclude our work.
\section{Background \& Related Works}
\label{sec:bg}

In this section we dive into the preliminary knowledge and the related works necessary to understand and place this work appropriately. 

\subsection{Privacy-Preserving Misbehavior Detection}

FL introduces unique challenges in handling malicious behavior due to its distributed nature. Byzantine attackers can disrupt training through two primary attack vectors: data poisoning and model poisoning. Data poisoning~\cite{fan2022survey} involves modifying local training data to mislead the global model. This includes strategies such as label flipping~\cite{jha2023label} (mislabeling training samples), perturbation~\cite{sandoval2022autoregressive} (injecting noise into data), and backdoor attacks~\cite{li2022backdoor} (embedding patterns in data that cause misclassifications). In contrast, model poisoning~\cite{wang2022defense} directly manipulates local model parameters to corrupt it, such as gradient inverting~\cite{wu2023learning} (maximizing the loss instead of minimizing it). 
Both attack types can be targeted (affecting specific classes or features) or untargeted (degrading model performance globally).  

Broadly, defense mechanisms against adversarial clients in FL fall into two main categories: mitigation and detection. Mitigation techniques aim to reduce the impact of malicious clients
by modifying the aggregation process itself. Common approaches include robust statistical methods such as  
metric-based comparisons using the Euclidean distance~\cite{blanchard2017machine} or cosine similarity~\cite{fung2018mitigating}. These methods assess local updates relative to each other or against a baseline (i.e., the global model). The works ~\cite{pillutla2022robust,BREA} propose robust aggregation techniques that mitigate the effect of poisoned models on the global model utility~\cite{pillutla2022robust,BREA}  without requiring access to individual local updates.

While mitigation strategies seek to minimize the influence of malicious clients, identifying which clients are malicious remains a crucial real-world problem~\cite{Kai21}. This is tackled by detection methods, which classify clients as honest or malicious, allowing the server to selectively exclude suspicious participants from training. While detection methods can effectively filter out harmful actors, they often rely on fine-grained client-level information, making them incompatible with privacy-preserving aggregation schemes such as SA. 

Privacy preservation and MD impose a natural trade-off because the former tries to hide the client's attributes (data, model updates, etc.), while the latter requires that the server learns more about the clients, such that a comparison is performed in an informative manner. However, both aspects of FL are important, and hence, it is crucial to balance this trade-off. 

\subsubsection*{FedGT} 

\begin{figure*}[!t]
    \centering
    \begin{subfigure}{0.3\textwidth}
        \resizebox{\textwidth}{!}{
        \begin{tikzpicture}[
    node distance=0.6cm and 1cm,
    basic/.style={draw, minimum size=0.8cm},
    basichexagon/.style={draw, minimum size=0.9cm},
    circle shape/.style={circle, basic},
    square shape/.style={rectangle, basic},
    triangle shape/.style={regular polygon, regular polygon sides=3, basic},
    hexagon shape/.style={regular polygon, regular polygon sides=6, basic},
    pentagon shape/.style={regular polygon, regular polygon sides=5, basichexagon}
]
\def\x{1.75cm}
\def\y{0.35cm}
\def\xshift{0.5cm}
\def\xshiftz{0.45cm}
\def\xshiftx{0.2cm}
\def\xshifty{0.25cm}
\def\xshiftyy{0.55cm}
\def\size{-5.2cm}
\def\sizex{-4.3cm}

\node[triangle shape, fill=gray!10] (a1) {};
\node[square shape, fill=gray!20, below=\y of a1] (a2) {};
\node[pentagon shape, fill=gray!20, below=\y of a2] (a3) {};
\node[hexagon shape, fill=gray!20, below=\y of a3] (a4) {};
\node[circle shape, fill=gray!20, below=\y of a4] (a5) {};

\node[triangle shape, fill=gray!40,right = \x of a1] (b1) {};
\node[square shape, fill=gray!40, below=\y of b1] (b2) {};
\node[pentagon shape, fill=gray!40, below=\y of b2] (b3) {};
\node[hexagon shape, fill=gray!40, below=\y of b3] (b4) {};
\node[circle shape, fill=gray!40, below=\y of b4] (b5) {};

\node[triangle shape, fill=gray!70,right = \x of b1] (c1) {};
\node[square shape, fill=gray!70, below=\y of c1] (c2) {};
\node[pentagon shape, fill=gray!70, below=\y of c2] (c3) {};
\node[hexagon shape, fill=gray!70, below=\y of c3] (c4) {};
\node[circle shape, fill=gray!70, below=\y of c4] (c5) {};

\node[triangle shape, fill=gray!70,right = \x of c1] (d1) {};
\node[square shape, fill=gray!100, below=\y of d1] (d2) {};
\node[pentagon shape, fill=gray!100, below=\y of d2] (d3) {};
\node[hexagon shape, fill=gray!100, below=\y of d3] (d4) {};
\node[circle shape, fill=gray!100, below=\y of d4] (d5) {};

\foreach \col/\topnode in {a/a1, b/b1, c/c1, d/d1} {
  \draw[line width=0.2pt,opacity=0.5]
    ([xshift=-0.2cm,yshift=.3cm]\topnode.north west) -- ++(-\xshiftx,0)
    -- ++(0,\size)
    -- ++(\xshiftx,0);
  \draw[line width=0.2pt,opacity=0.5]
    ([xshift=0.2cm,yshift=.3cm]\topnode.north east) -- ++(\xshiftx,0)
    -- ++(0,\size)
    -- ++(-\xshiftx,0);
}

\foreach \col/\topnode in {a/a1, b/b1, c/c1} {
    \draw[line width=0.3pt,opacity=0.75]   ([xshift=\xshift,yshift=-.15cm]\topnode.north east) -- ++(\xshiftx,0)
    -- ++(0,\sizex)
    -- ++(-\xshiftx,0);
}

\foreach \col/\topnode in {b/b1, c/c1, d/d1} {
  \draw[line width=0.3pt,opacity=0.75]([xshift=-\xshift,yshift=-.15cm]\topnode.north west) -- ++(-\xshiftx,0)
    -- ++(0,\sizex)
    -- ++(\xshiftx,0);
}

\foreach \i in {2,3,4} {
  \draw[solid, opacity = 0.75]
    ($(a\i.east) + (\xshifty,0)$) -- ++(.2cm,0);
    \draw[solid, opacity = 0.75]    ($(b\i.east) + (\xshifty,0)$) -- ++(.2cm,0);
    \draw[solid, opacity = 0.75]    ($(c\i.east) + (\xshifty,0)$) -- ++(.2cm,0);
    \draw[solid, opacity = 0.75]
    ($(b\i.west) + (-\xshifty,0)$) -- ++(-.2cm,0);
    \draw[solid, opacity = 0.75]
    ($(c\i.west) + (-\xshifty,0)$) -- ++(-.2cm,0);
    \draw[solid, opacity = 0.75]
    ($(d\i.west) + (-\xshifty,0)$) -- ++(-.2cm,0);
}

\draw[solid,ultra thick,opacity = 0.75]
    ($(a3.east) + (\xshiftz,0)$) -- ++ (\xshiftyy,0);
\draw[solid,ultra thick,opacity = 0.75]
    ($(b3.east) + (\xshiftz,0)$) -- ++ (\xshiftyy,0);
\draw[solid,ultra thick,opacity = 0.75]
    ($(c3.east) + (\xshiftz,0)$) -- ++ (\xshiftyy,0);

\begin{scope}[xshift=0, yshift=-7cm, every node/.style={scale=1}]
\node[] (matrix) at (1,0.2) {};    
\def\scale{0.72}
\node[triangle shape, fill=gray!10, below right = -2.5 and 1 of a1] (T1) at (1, -1) {}; 
\node[pentagon shape, fill=gray!10, below right = -.47 and .3cm of T1,scale=\scale] (P1) {};
\node[hexagon shape, fill=gray!10, right = .15cm of P1,scale = \scale] (H1) {};
\node[square shape, fill=gray!40, below =0.2 of T1,scale=\scale] (S2) {};
\node[pentagon shape, fill=gray!40, below = 0.2 of P1,scale=\scale] (P2) {};
\node[circle shape, fill=gray!40, below = 0.2 of H1,scale=\scale] (C2) {};
\node[triangle shape, fill=gray!70, below =0.2 of S2] (T3) {};
\node[pentagon shape, fill=gray!70, below = 0.2 of P2,scale=\scale] (P3) {};
\node[circle shape, fill=gray!70, below = 0.2 of C2,scale=\scale] (C3) {};

\draw[line width=0.5pt, opacity = 0.7]
  ($(T1.west)+(0,0.45)$) -- ++(-0.2, 0) -- ++(0, -.75) -- ++(0.2, 0); 
\draw[line width=0.5pt, opacity = 0.7]
  ($(S2.west)+(0.05,0.35)$) -- ++(-0.2, 0) -- ++(0, -.7) -- ++(0.2, 0);
\draw[line width=0.5pt, opacity = 0.7]
  ($(T3.west)+(0,0.5)$) -- ++(-0.2, 0) -- ++(0, -.75) -- ++(0.2, 0); 

\draw[line width=0.5pt, opacity = 0.7]
  ($(H1.east)+(-0.13,0.4)$) -- ++(0.2, 0) -- ++(0, -.75) -- ++(-0.2, 0);
\draw[line width=0.5pt, opacity = 0.7]
  ($(C2.east)+(-0.13,0.35)$) -- ++(0.2, 0) -- ++(0, -.7) -- ++(-0.2, 0);
\draw[line width=0.5pt, opacity = 0.7]
  ($(C3.east)+(-0.13,0.35)$) -- ++(0.2, 0) -- ++(0, -.7) -- ++(-0.2, 0);

\node[below right = 0 and .7 of H1] (Compare1) {\footnotesize Compare};
\node[below = .3 of Compare1] (Compare2) {\footnotesize Compare};

\node[left = .35 of T1.west] (round2) {\footnotesize Round $1$};
\node[left = .3 of S2.west] (round2) {\footnotesize Round $2$};
\node[above left = 0 and .35 of T3.west] (round2) {\footnotesize Round $3$};

\draw[line width = 0.5pt, opacity = 0.9, ->]
($(H1.east) + (.1,0)$) -- (Compare1.west);
\draw[line width = 0.5pt, opacity = 0.9, ->]
($(C2.east) + (.1,0)$) -- (Compare1.west);

\draw[line width = 0.5pt, opacity = 0.9, ->]
($(C2.east) + (.1,-.1)$) -- (Compare2.west);
\draw[line width = 0.5pt, opacity = 0.9, ->]
($(C3.east) + (.1,0)$) -- (Compare2.west);

\end{scope}
\end{tikzpicture}
        }
        \caption{\footnotesize Cross-Device QI.}
        \label{fig:dev-qi}
    \end{subfigure}
    \hfill
    \begin{subfigure}{0.3\textwidth}
        \resizebox{\textwidth}{!}{
        \begin{tikzpicture}[
    node distance=0.6cm and 1cm,
    basic/.style={draw, minimum size=0.8cm},
    basichexagon/.style={draw, minimum size=0.9cm},
    circle shape/.style={circle, basic},
    square shape/.style={rectangle, basic},
    triangle shape/.style={regular polygon, regular polygon sides=3, basic},
    hexagon shape/.style={regular polygon, regular polygon sides=6, basic},
    pentagon shape/.style={regular polygon, regular polygon sides=5, basichexagon}
]
\def\x{1.75cm}
\def\y{0.35cm}
\def\xshift{0.5cm}
\def\xshiftz{0.45cm}
\def\xshiftx{0.2cm}
\def\xshifty{0.25cm}
\def\xshiftyy{0.55cm}
\def\size{-5.2cm}
\def\sizex{-4.3cm}

\node[triangle shape, fill=gray!10] (a1) {};
\node[square shape, fill=gray!20, below=\y of a1] (a2) {};
\node[pentagon shape, fill=gray!20, below=\y of a2] (a3) {};
\node[hexagon shape, fill=gray!20, below=\y of a3] (a4) {};
\node[circle shape, fill=gray!20, below=\y of a4] (a5) {};

\node[triangle shape, fill=gray!40,right = \x of a1] (b1) {};
\node[square shape, fill=gray!40, below=\y of b1] (b2) {};
\node[pentagon shape, fill=gray!40, below=\y of b2] (b3) {};
\node[hexagon shape, fill=gray!40, below=\y of b3] (b4) {};
\node[circle shape, fill=gray!40, below=\y of b4] (b5) {};

\node[triangle shape, fill=gray!70,right = \x of b1] (c1) {};
\node[square shape, fill=gray!70, below=\y of c1] (c2) {};
\node[pentagon shape, fill=gray!70, below=\y of c2] (c3) {};
\node[hexagon shape, fill=gray!70, below=\y of c3] (c4) {};
\node[circle shape, fill=gray!70, below=\y of c4] (c5) {};

\node[triangle shape, fill=gray!70,right = \x of c1] (d1) {};
\node[square shape, fill=gray!100, below=\y of d1] (d2) {};
\node[pentagon shape, fill=gray!100, below=\y of d2] (d3) {};
\node[hexagon shape, fill=gray!100, below=\y of d3] (d4) {};
\node[circle shape, fill=gray!100, below=\y of d4] (d5) {};

\foreach \col/\topnode in {a/a1, b/b1, c/c1, d/d1} {
  \draw[line width=0.2pt,opacity=0.5]
    ([xshift=-0.2cm,yshift=.3cm]\topnode.north west) -- ++(-\xshiftx,0)
    -- ++(0,\size)
    -- ++(\xshiftx,0);
  \draw[line width=0.2pt,opacity=0.5]
    ([xshift=0.2cm,yshift=.3cm]\topnode.north east) -- ++(\xshiftx,0)
    -- ++(0,\size)
    -- ++(-\xshiftx,0);
}

\foreach \col/\topnode in {a/a1, b/b1, c/c1} {
    \draw[line width=0.3pt,opacity=0.75]   ([xshift=\xshift,yshift=-.15cm]\topnode.north east) -- ++(\xshiftx,0)
    -- ++(0,\sizex)
    -- ++(-\xshiftx,0);
}

\foreach \col/\topnode in {b/b1, c/c1, d/d1} {
  \draw[line width=0.3pt,opacity=0.75]([xshift=-\xshift,yshift=-.15cm]\topnode.north west) -- ++(-\xshiftx,0)
    -- ++(0,\sizex)
    -- ++(\xshiftx,0);
}

\foreach \i in {2,3,4} {
  \draw[solid, opacity = 0.75]
    ($(a\i.east) + (\xshifty,0)$) -- ++(.2cm,0);
    \draw[solid, opacity = 0.75]    ($(b\i.east) + (\xshifty,0)$) -- ++(.2cm,0);
    \draw[solid, opacity = 0.75]    ($(c\i.east) + (\xshifty,0)$) -- ++(.2cm,0);
    \draw[solid, opacity = 0.75]
    ($(b\i.west) + (-\xshifty,0)$) -- ++(-.2cm,0);
    \draw[solid, opacity = 0.75]
    ($(c\i.west) + (-\xshifty,0)$) -- ++(-.2cm,0);
    \draw[solid, opacity = 0.75]
    ($(d\i.west) + (-\xshifty,0)$) -- ++(-.2cm,0);
}

\draw[solid,ultra thick,opacity = 0.75]
    ($(a3.east) + (\xshiftz,0)$) -- ++ (\xshiftyy,0);
\draw[solid,ultra thick,opacity = 0.75]
    ($(b3.east) + (\xshiftz,0)$) -- ++ (\xshiftyy,0);
\draw[solid,ultra thick,opacity = 0.75]
    ($(c3.east) + (\xshiftz,0)$) -- ++ (\xshiftyy,0);

\begin{scope}[xshift=0, yshift=-7cm, every node/.style={scale=1}]
\node[] (matrix) at (1,0.2) {};    
\def\scale{0.72}
\node[triangle shape, fill=gray!40, below right = -2.5 and 1 of a1] (T1) at (1, -1) {}; 
\node[pentagon shape, fill=gray!40, below right = -.47 and .3cm of T1,scale=\scale] (P1) {};
\node[hexagon shape, fill=gray!40, right = .15cm of P1,scale = \scale] (H1) {};
\node[square shape, fill=gray!40, below =0.2 of T1,scale=\scale] (S2) {};
\node[pentagon shape, fill=gray!40, below = 0.2 of P1,scale=\scale] (P2) {};
\node[circle shape, fill=gray!40, below = 0.2 of H1,scale=\scale] (C2) {};
\node[triangle shape, fill=gray!40, below =0.2 of S2] (T3) {};
\node[pentagon shape, fill=gray!40, below = 0.2 of P2,scale=\scale] (P3) {};
\node[circle shape, fill=gray!40, below = 0.2 of C2,scale=\scale] (C3) {};

\draw[line width=0.5pt, opacity = 0.7]
  ($(T1.west)+(0,0.45)$) -- ++(-0.2, 0) -- ++(0, -.75) -- ++(0.2, 0); 
\draw[line width=0.5pt, opacity = 0.7]
  ($(S2.west)+(0.05,0.35)$) -- ++(-0.2, 0) -- ++(0, -.7) -- ++(0.2, 0);
\draw[line width=0.5pt, opacity = 0.7]
  ($(T3.west)+(0,0.5)$) -- ++(-0.2, 0) -- ++(0, -.75) -- ++(0.2, 0); 

\draw[line width=0.5pt, opacity = 0.7]
  ($(H1.east)+(-0.13,0.4)$) -- ++(0.2, 0) -- ++(0, -.75) -- ++(-0.2, 0);
\draw[line width=0.5pt, opacity = 0.7]
  ($(C2.east)+(-0.13,0.35)$) -- ++(0.2, 0) -- ++(0, -.7) -- ++(-0.2, 0);
\draw[line width=0.5pt, opacity = 0.7]
  ($(C3.east)+(-0.13,0.35)$) -- ++(0.2, 0) -- ++(0, -.7) -- ++(-0.2, 0);

\node[below right = -.2 and .7 of H1] (Compare1) {\footnotesize Compare};
\node[below = .2 of Compare1] (Compare2) {\footnotesize Compare};
\node[below = .2 of Compare2] (Compare3) {\footnotesize Compare};

\node[left = .35 of T1.west] (round2) {\footnotesize Round $2$};
\node[above left = 0 and .35 of T3.west] (round2) {\footnotesize Round $2$};
\node[left = .3 of S2.west] (round2) {\footnotesize Round $2$};

\draw[line width = 0.5pt, opacity = 0.9, ->]
($(H1.east) + (.1,0)$) -- (Compare1.west);
\draw[line width = 0.5pt, opacity = 0.9, ->]
($(C2.east) + (.1,0)$) -- (Compare1.west);

\draw[line width = 0.5pt, opacity = 0.9, ->]
($(H1.east) + (.1,0)$) -- (Compare2.west);
\draw[line width = 0.5pt, opacity = 0.9, ->]
($(C3.east) + (.1,0)$) -- (Compare2.west);

\draw[line width = 0.5pt, opacity = 0.9, ->]
($(C2.east) + (.1,0)$) -- (Compare3.west);
\draw[line width = 0.5pt, opacity = 0.9, ->]
($(C3.east) + (.1,0)$) -- (Compare3.west);

\end{scope}
\end{tikzpicture}
        }
        \caption{\footnotesize Cross-Silo QI.}
        \label{fig:sil-qi}
    \end{subfigure}
    \hfill
    \begin{subfigure}{0.3\textwidth}
        \includegraphics[width=\linewidth]{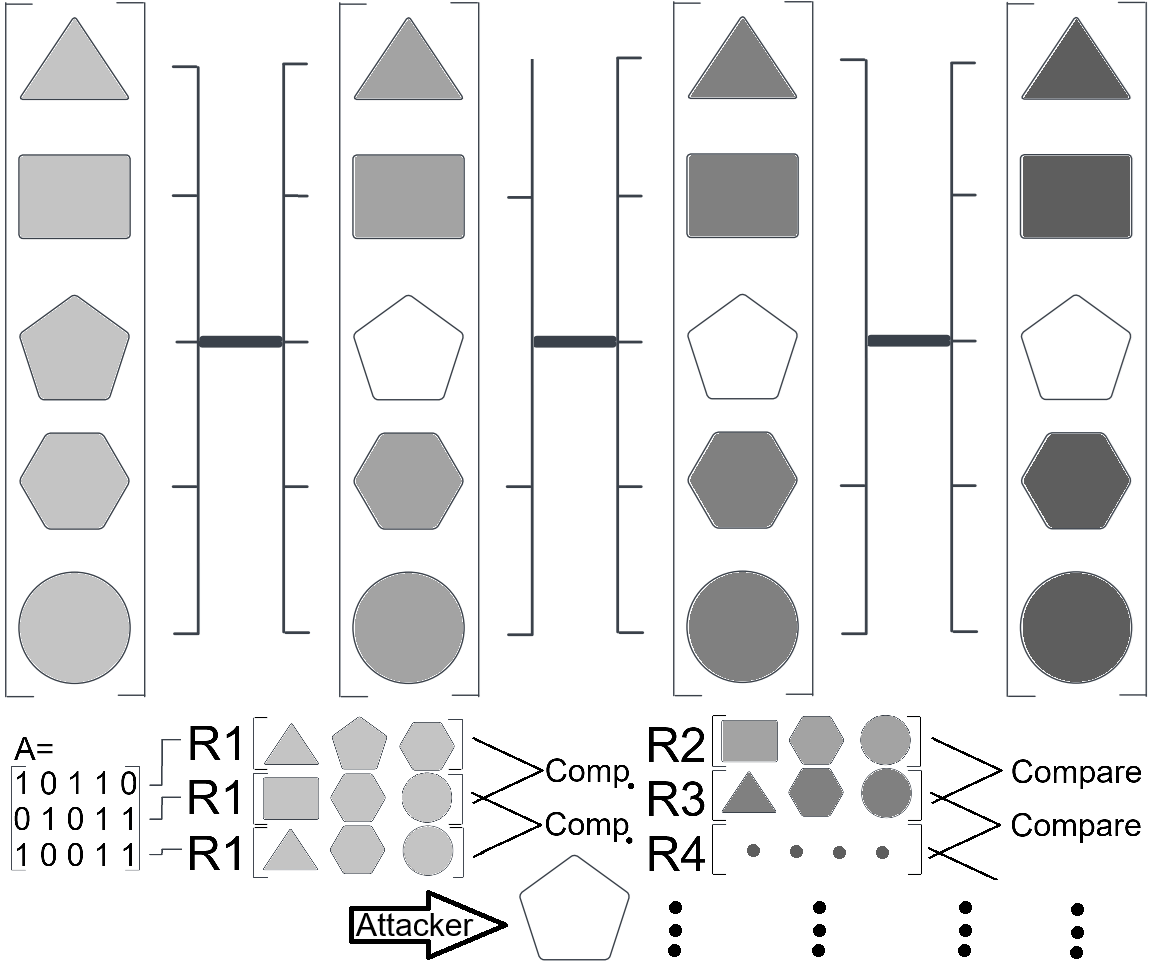}
        \caption{\footnotesize QI with within and across round testing.}
        \label{fig:dev-comb}
    \end{subfigure}
    \hfill
    \begin{subfigure}{0.3\textwidth}
        \includegraphics[width=\linewidth]{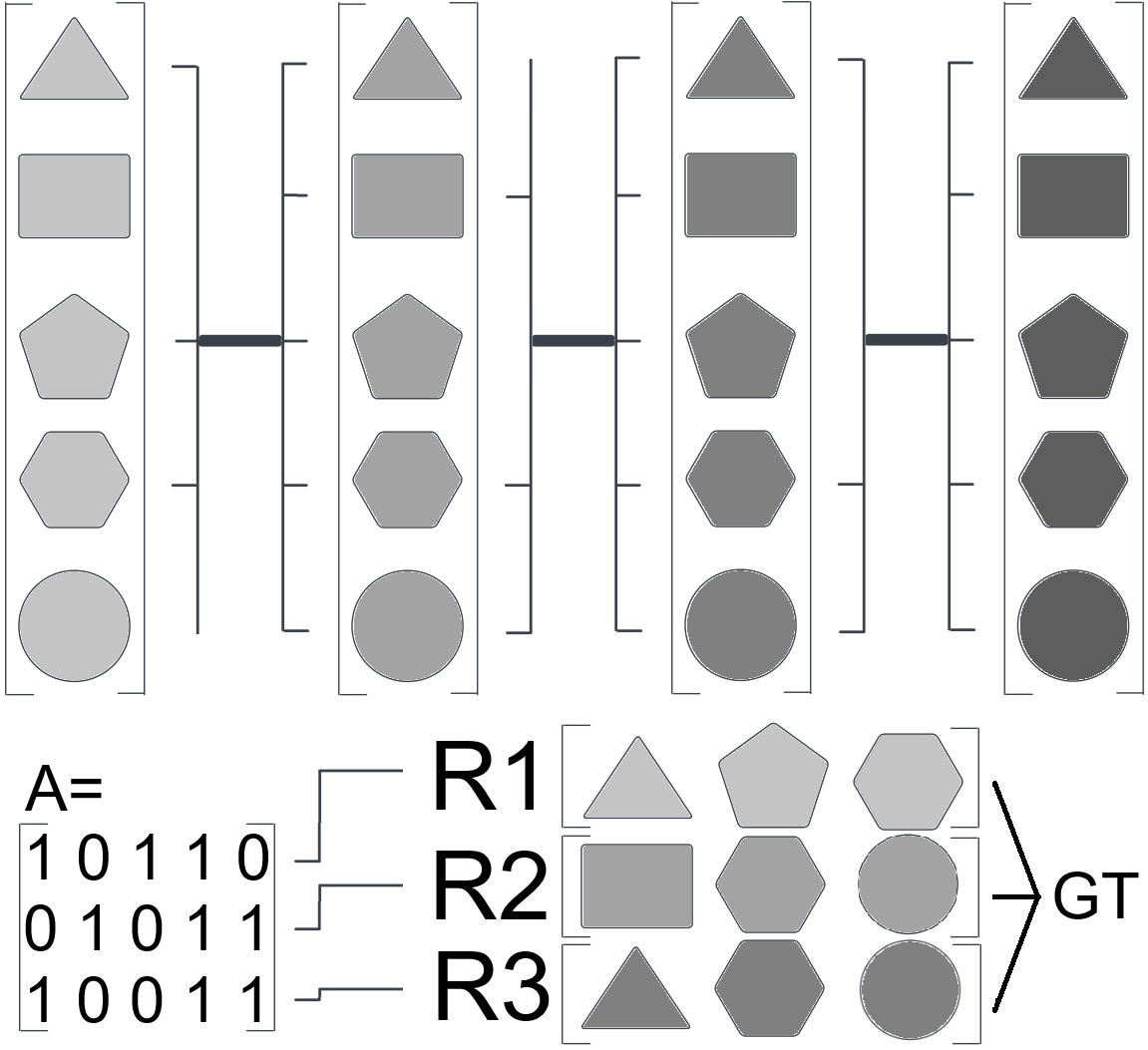}
        \caption{\footnotesize Cross-Device FedGT.}
        \label{fig:dev-gt}
    \end{subfigure}
    \hfill
    \begin{subfigure}{0.3\textwidth}
        \resizebox{\textwidth}{!}{
        \begin{tikzpicture}[
    node distance=0.6cm and 1cm,
    basic/.style={draw, minimum size=0.8cm},
    basichexagon/.style={draw, minimum size=0.9cm},
    circle shape/.style={circle, basic},
    square shape/.style={rectangle, basic},
    triangle shape/.style={regular polygon, regular polygon sides=3, basic},
    hexagon shape/.style={regular polygon, regular polygon sides=6, basic},
    pentagon shape/.style={regular polygon, regular polygon sides=5, basichexagon}
]
\def\x{1.75cm}
\def\y{0.35cm}
\def\xshift{0.5cm}
\def\xshiftz{0.45cm}
\def\xshiftx{0.2cm}
\def\xshifty{0.25cm}
\def\xshiftyy{0.55cm}
\def\size{-5.2cm}
\def\sizex{-4.3cm}

\node[triangle shape, fill=gray!10] (a1) {};
\node[square shape, fill=gray!20, below=\y of a1] (a2) {};
\node[pentagon shape, fill=gray!20, below=\y of a2] (a3) {};
\node[hexagon shape, fill=gray!20, below=\y of a3] (a4) {};
\node[circle shape, fill=gray!20, below=\y of a4] (a5) {};

\node[triangle shape, fill=gray!40,right = \x of a1] (b1) {};
\node[square shape, fill=gray!40, below=\y of b1] (b2) {};
\node[pentagon shape, fill=gray!40, below=\y of b2] (b3) {};
\node[hexagon shape, fill=gray!40, below=\y of b3] (b4) {};
\node[circle shape, fill=gray!40, below=\y of b4] (b5) {};

\node[triangle shape, fill=gray!70,right = \x of b1] (c1) {};
\node[square shape, fill=gray!70, below=\y of c1] (c2) {};
\node[pentagon shape, fill=gray!70, below=\y of c2] (c3) {};
\node[hexagon shape, fill=gray!70, below=\y of c3] (c4) {};
\node[circle shape, fill=gray!70, below=\y of c4] (c5) {};

\node[triangle shape, fill=gray!100,right = \x of c1] (d1) {};
\node[square shape, fill=gray!100, below=\y of d1] (d2) {};
\node[pentagon shape, fill=gray!100, below=\y of d2] (d3) {};
\node[hexagon shape, fill=gray!100, below=\y of d3] (d4) {};
\node[circle shape, fill=gray!100, below=\y of d4] (d5) {};

\foreach \col/\topnode in {a/a1, b/b1, c/c1, d/d1} {
  \draw[line width=0.2pt,opacity=0.5]
    ([xshift=-0.2cm,yshift=.3cm]\topnode.north west) -- ++(-\xshiftx,0)
    -- ++(0,\size)
    -- ++(\xshiftx,0);
  \draw[line width=0.2pt,opacity=0.5]
    ([xshift=0.2cm,yshift=.3cm]\topnode.north east) -- ++(\xshiftx,0)
    -- ++(0,\size)
    -- ++(-\xshiftx,0);
}

\foreach \col/\topnode in {a/a1, b/b1, c/c1} {
    \draw[line width=0.3pt,opacity=0.75]   ([xshift=\xshift,yshift=-.15cm]\topnode.north east) -- ++(\xshiftx,0)
    -- ++(0,\sizex)
    -- ++(-\xshiftx,0);
}

\foreach \col/\topnode in {b/b1, c/c1, d/d1} {
  \draw[line width=0.3pt,opacity=0.75]([xshift=-\xshift,yshift=-.15cm]\topnode.north west) -- ++(-\xshiftx,0)
    -- ++(0,\sizex)
    -- ++(\xshiftx,0);
}

\foreach \i in {2,3,4} {
  \draw[solid, opacity = 0.75]
    ($(a\i.east) + (\xshifty,0)$) -- ++(.2cm,0);
    \draw[solid, opacity = 0.75]    ($(b\i.east) + (\xshifty,0)$) -- ++(.2cm,0);
    \draw[solid, opacity = 0.75]    ($(c\i.east) + (\xshifty,0)$) -- ++(.2cm,0);
    \draw[solid, opacity = 0.75]
    ($(b\i.west) + (-\xshifty,0)$) -- ++(-.2cm,0);
    \draw[solid, opacity = 0.75]
    ($(c\i.west) + (-\xshifty,0)$) -- ++(-.2cm,0);
    \draw[solid, opacity = 0.75]
    ($(d\i.west) + (-\xshifty,0)$) -- ++(-.2cm,0);
}

\draw[solid,ultra thick,opacity = 0.75]
    ($(a3.east) + (\xshiftz,0)$) -- ++ (\xshiftyy,0);
\draw[solid,ultra thick,opacity = 0.75]
    ($(b3.east) + (\xshiftz,0)$) -- ++ (\xshiftyy,0);
\draw[solid,ultra thick,opacity = 0.75]
    ($(c3.east) + (\xshiftz,0)$) -- ++ (\xshiftyy,0);

\begin{scope}[xshift=0, yshift=-7cm, every node/.style={scale=1}]
\node[] (matrix) at (1,0.2) {
    $A=
    \begin{pmatrix}
    1 & 0 & 1 & 1 & 0 \\
    0 & 1 & 0 & 1 & 1 \\
    1 & 0 & 0 & 1 & 1 \\
    \end{pmatrix}$
};    
\def\scale{0.72}
\node[triangle shape, fill=gray!40, below right = -2.5 and 3 of a1] (T1) at (1, -1) {}; 
\node[pentagon shape, fill=gray!40, below right = -.47 and .3cm of T1,scale=\scale] (P1) {};
\node[hexagon shape, fill=gray!40, right = .15cm of P1,scale = \scale] (H1) {};
\node[square shape, fill=gray!40, below =0.2 of T1,scale=\scale] (S2) {};
\node[pentagon shape, fill=gray!40, below = 0.2 of P1,scale=\scale] (P2) {};
\node[circle shape, fill=gray!40, below = 0.2 of H1,scale=\scale] (C2) {};
\node[triangle shape, fill=gray!40, below =0.2 of S2] (T3) {};
\node[pentagon shape, fill=gray!40, below = 0.2 of P2,scale=\scale] (P3) {};
\node[circle shape, fill=gray!40, below = 0.2 of C2,scale=\scale] (C3) {};

\draw[line width=0.5pt, opacity = 0.7]
  ($(T1.west)+(0,0.45)$) -- ++(-0.2, 0) -- ++(0, -.75) -- ++(0.2, 0); 
\draw[line width=0.5pt, opacity = 0.7]
  ($(S2.west)+(0.05,0.35)$) -- ++(-0.2, 0) -- ++(0, -.7) -- ++(0.2, 0);
\draw[line width=0.5pt, opacity = 0.7]
  ($(T3.west)+(0,0.5)$) -- ++(-0.2, 0) -- ++(0, -.75) -- ++(0.2, 0); 

\draw[line width=0.5pt, opacity = 0.7]
  ($(H1.east)+(-0.13,0.4)$) -- ++(0.2, 0) -- ++(0, -.75) -- ++(-0.2, 0);
\draw[line width=0.5pt, opacity = 0.7]
  ($(C2.east)+(-0.13,0.35)$) -- ++(0.2, 0) -- ++(0, -.7) -- ++(-0.2, 0);
\draw[line width=0.5pt, opacity = 0.7]
  ($(C3.east)+(-0.13,0.35)$) -- ++(0.2, 0) -- ++(0, -.7) -- ++(-0.2, 0);

\node[right = .3 of C2] (FedGT) {\footnotesize FedGT};

\node[left = .35 of T1.west,rotate=90] (round2) {\footnotesize Round $2$};

\draw[line width = 0.3pt, opacity = 0.5, ->]
($(H1.east) + (.1,0)$) -| (FedGT);

\draw[line width = 0.3pt, opacity = 0.5, ->]
($(C2.east) + (.1,0)$) -- (FedGT.west);

\draw[line width = 0.3pt, opacity = 0.5, ->]
($(C3.east) + (.1,0)$) -| (FedGT);

\draw[line width = 0.5pt, opacity = 0.9, ->]
($(matrix.east) + (0, .4)$) -- ++ (0.2,0) -- ++ (0, .8) -> ++(.9, 0);
\draw[line width = 0.5pt, opacity = 0.9, ->]
($(matrix.east) + (0, 0)$) -- ++ (0.4,0) -- ++ (0, .4) -> ++(.4, 0);
\draw[line width = 0.5pt, opacity = 0.9, ->]
($(matrix.east) + (0, -.4)$) -> ++(1.1, 0);

\end{scope}
\end{tikzpicture}
        }
        \caption{\footnotesize Cross-Silo FedGT.}
        \label{fig:sil-gt}
    \end{subfigure}
    \hfill
    \begin{subfigure}{0.3\textwidth}
        \centering
        \resizebox{\textwidth}{!}{\begin{tikzpicture}[
    node distance=0.6cm and 1cm,
    basic/.style={draw, minimum size=0.8cm},
    basichexagon/.style={draw, minimum size=0.9cm},
    circle shape/.style={circle, basic},
    square shape/.style={rectangle, basic},
    triangle shape/.style={regular polygon, regular polygon sides=3, basic},
    hexagon shape/.style={regular polygon, regular polygon sides=6, basic},
    pentagon shape/.style={regular polygon, regular polygon sides=5, basichexagon}
]
\def\x{1.75cm}
\def\y{0.35cm}
\def\xshift{0.5cm}
\def\xshiftz{0.45cm}
\def\xshiftx{0.2cm}
\def\xshifty{0.25cm}
\def\xshiftyy{0.55cm}
\def\size{-5.2cm}
\def\sizex{-4.3cm}

\node[triangle shape, fill=gray!10] (a1) {};
\node[square shape, fill=gray!20, below=\y of a1] (a2) {};
\node[pentagon shape, fill=gray!20, below=\y of a2] (a3) {};
\node[hexagon shape, fill=gray!20, below=\y of a3] (a4) {};
\node[circle shape, fill=gray!20, below=\y of a4] (a5) {};

\node[triangle shape, fill=gray!40,right = \x of a1] (b1) {};
\node[square shape, fill=gray!40, below=\y of b1] (b2) {};
\node[pentagon shape, fill=gray!40, below=\y of b2] (b3) {};
\node[hexagon shape, fill=gray!40, below=\y of b3] (b4) {};
\node[circle shape, fill=gray!40, below=\y of b4] (b5) {};

\node[triangle shape, fill=gray!70,right = \x of b1] (c1) {};
\node[square shape, fill=gray!70, below=\y of c1] (c2) {};
\node[pentagon shape, fill=gray!70, below=\y of c2] (c3) {};
\node[hexagon shape, fill=gray!70, below=\y of c3] (c4) {};
\node[circle shape, fill=gray!70, below=\y of c4] (c5) {};

\node[triangle shape, fill=gray!70,right = \x of c1] (d1) {};
\node[square shape, fill=gray!100, below=\y of d1] (d2) {};
\node[pentagon shape, fill=gray!100, below=\y of d2] (d3) {};
\node[hexagon shape, fill=gray!100, below=\y of d3] (d4) {};
\node[circle shape, fill=gray!100, below=\y of d4] (d5) {};

\foreach \col/\topnode in {a/a1, b/b1, c/c1, d/d1} {
  \draw[line width=0.2pt,opacity=0.5]
    ([xshift=-0.2cm,yshift=.3cm]\topnode.north west) -- ++(-\xshiftx,0)
    -- ++(0,\size)
    -- ++(\xshiftx,0);
  \draw[line width=0.2pt,opacity=0.5]
    ([xshift=0.2cm,yshift=.3cm]\topnode.north east) -- ++(\xshiftx,0)
    -- ++(0,\size)
    -- ++(-\xshiftx,0);
}

\foreach \col/\topnode in {a/a1, b/b1, c/c1} {
    \draw[line width=0.3pt,opacity=0.75]   ([xshift=\xshift,yshift=-.15cm]\topnode.north east) -- ++(\xshiftx,0)
    -- ++(0,\sizex)
    -- ++(-\xshiftx,0);
}

\foreach \col/\topnode in {b/b1, c/c1, d/d1} {
  \draw[line width=0.3pt,opacity=0.75]([xshift=-\xshift,yshift=-.15cm]\topnode.north west) -- ++(-\xshiftx,0)
    -- ++(0,\sizex)
    -- ++(\xshiftx,0);
}

\foreach \i in {2,3,4} {
  \draw[solid, opacity = 0.75]
    ($(a\i.east) + (\xshifty,0)$) -- ++(.2cm,0);
    \draw[solid, opacity = 0.75]    ($(b\i.east) + (\xshifty,0)$) -- ++(.2cm,0);
    \draw[solid, opacity = 0.75]    ($(c\i.east) + (\xshifty,0)$) -- ++(.2cm,0);
    \draw[solid, opacity = 0.75]
    ($(b\i.west) + (-\xshifty,0)$) -- ++(-.2cm,0);
    \draw[solid, opacity = 0.75]
    ($(c\i.west) + (-\xshifty,0)$) -- ++(-.2cm,0);
    \draw[solid, opacity = 0.75]
    ($(d\i.west) + (-\xshifty,0)$) -- ++(-.2cm,0);
}

\draw[solid,ultra thick,opacity = 0.75]
    ($(a3.east) + (\xshiftz,0)$) -- ++ (\xshiftyy,0);
\draw[solid,ultra thick,opacity = 0.75]
    ($(b3.east) + (\xshiftz,0)$) -- ++ (\xshiftyy,0);
\draw[solid,ultra thick,opacity = 0.75]
    ($(c3.east) + (\xshiftz,0)$) -- ++ (\xshiftyy,0);

\begin{scope}[xshift=0, yshift=-7cm, every node/.style={scale=1}]
\node[] (matrix) at (3,1.5) {
    $A=
    \begin{pmatrix}
    1 & 0 & 1 & 1 & 0 \\
    0 & 1 & 0 & 1 & 1 \\
    1 & 0 & 0 & 1 & 1 \\
    \end{pmatrix}$
};    
\def\scale{0.72}
\node[triangle shape, fill=gray!40, above left = 1.5 and 1.5cm of matrix] (T1) at (1, -1) {}; 
\node[pentagon shape, fill=gray!40, below right = -.47 and .3cm of T1,scale=\scale] (P1) {};
\node[hexagon shape, fill=gray!40, right = .15cm of P1,scale = \scale] (H1) {};
\node[square shape, fill=gray!40, below =0.2 of T1,scale=\scale] (S2) {};
\node[pentagon shape, fill=gray!40, below = 0.2 of P1,scale=\scale] (P2) {};
\node[circle shape, fill=gray!40, below = 0.2 of H1,scale=\scale] (C2) {};
\node[triangle shape, fill=gray!40, below =0.2 of S2] (T3) {};
\node[pentagon shape, fill=gray!40, below = 0.2 of P2,scale=\scale] (P3) {};
\node[circle shape, fill=gray!40, below = 0.2 of C2,scale=\scale] (C3) {};

\node[triangle shape, fill=gray!70, right = 5.5 of T1] (T31) {};
\node[pentagon shape, fill=gray!70, below right = -.47 and .3cm of T31,scale=\scale] (P31) {};
\node[hexagon shape, fill=gray!70, right = .15cm of P31,scale = \scale] (H31) {};
\node[square shape, fill=gray!70, below =0.2 of T31,scale=\scale] (S32) {};
\node[pentagon shape, fill=gray!70, below = 0.2 of P31,scale=\scale] (P32) {};
\node[circle shape, fill=gray!70, below = 0.2 of H31,scale=\scale] (C32) {};
\node[triangle shape, fill=gray!70, below =0.2 of S32] (T33) {};
\node[pentagon shape, fill=gray!70, below = 0.2 of P32,scale=\scale] (P33) {};
\node[circle shape, fill=gray!70, below = 0.2 of C32,scale=\scale] (C33) {};

\draw[line width=0.5pt, opacity = 0.7]
  ($(T1.west)+(0,0.45)$) -- ++(-0.2, 0) -- ++(0, -.75) -- ++(0.2, 0); 
\draw[line width=0.5pt, opacity = 0.7]
  ($(S2.west)+(0.05,0.35)$) -- ++(-0.2, 0) -- ++(0, -.7) -- ++(0.2, 0);
\draw[line width=0.5pt, opacity = 0.7]
  ($(T3.west)+(0,0.5)$) -- ++(-0.2, 0) -- ++(0, -.75) -- ++(0.2, 0); 

\draw[line width=0.5pt, opacity = 0.7]
  ($(T31.west)+(0,0.45)$) -- ++(-0.2, 0) -- ++(0, -.75) -- ++(0.2, 0); 
\draw[line width=0.5pt, opacity = 0.7]
  ($(S32.west)+(0.05,0.35)$) -- ++(-0.2, 0) -- ++(0, -.7) -- ++(0.2, 0);
\draw[line width=0.5pt, opacity = 0.7]
  ($(T33.west)+(0,0.5)$) -- ++(-0.2, 0) -- ++(0, -.75) -- ++(0.2, 0); 

\draw[line width=0.5pt, opacity = 0.7]
  ($(H1.east)+(-0.13,0.4)$) -- ++(0.2, 0) -- ++(0, -.75) -- ++(-0.2, 0);
\draw[line width=0.5pt, opacity = 0.7]
  ($(C2.east)+(-0.13,0.35)$) -- ++(0.2, 0) -- ++(0, -.7) -- ++(-0.2, 0);
\draw[line width=0.5pt, opacity = 0.7]
  ($(C3.east)+(-0.13,0.35)$) -- ++(0.2, 0) -- ++(0, -.7) -- ++(-0.2, 0);

\draw[line width=0.5pt, opacity = 0.7]
  ($(H31.east)+(-0.13,0.4)$) -- ++(0.2, 0) -- ++(0, -.75) -- ++(-0.2, 0);
\draw[line width=0.5pt, opacity = 0.7]
  ($(C32.east)+(-0.13,0.35)$) -- ++(0.2, 0) -- ++(0, -.7) -- ++(-0.2, 0);
\draw[line width=0.5pt, opacity = 0.7]
  ($(C33.east)+(-0.13,0.35)$) -- ++(0.2, 0) -- ++(0, -.7) -- ++(-0.2, 0);

\node[right = .3 of C2] (FedGT) {\footnotesize FedGT};

\node[left = .35 of T1.west,rotate=90] (round2) {\footnotesize Round $2$};
\node[above right = 0.5 and .1 of C32.east,rotate=270] (round3) {\footnotesize Round $3$};

\node[left = .4 of S32] (FedGT3) {\footnotesize FedGT};

\draw[line width = 0.3pt, opacity = 0.5, ->]
($(H1.east) + (.1,0)$) -| (FedGT);

\draw[line width = 0.3pt, opacity = 0.5, ->]
($(C2.east) + (.1,0)$) -- (FedGT.west);

\draw[line width = 0.3pt, opacity = 0.5, ->]
($(C3.east) + (.1,0)$) -| (FedGT);

\draw[line width = 0.3pt, opacity = 0.5, ->]
($(T31.west) + (-.2,0)$) -| (FedGT3);
\draw[line width = 0.3pt, opacity = 0.5, ->]
($(T33.west) + (-.2,0)$) -| (FedGT3);
\draw[line width = 0.3pt, opacity = 0.5, ->]
($(S32.west) + (-.2,0)$) -- (FedGT3.east);

\node[below left = .5 and -.3 of FedGT3] (MRGT) {\footnotesize Multi-Round FedGT};

\draw[line width = 0.5pt, opacity = 0.75, ->]
(FedGT.east) -| (MRGT.north);
\draw[line width = 0.5pt, opacity = 0.75, ->]
($(FedGT3.west) + (0,.023)$) -| (MRGT.north);
  
\end{scope}
\end{tikzpicture}}
        \caption{\footnotesize Cross-Silo Multi-Round FedGT.}
        \label{fig:sil-multi}
    \end{subfigure}
    \caption{Illustration of the envisioned settings where the shapes represent clients and the grayness represents the rounds. 
    }
    \label{fig:illustrate}
\end{figure*}
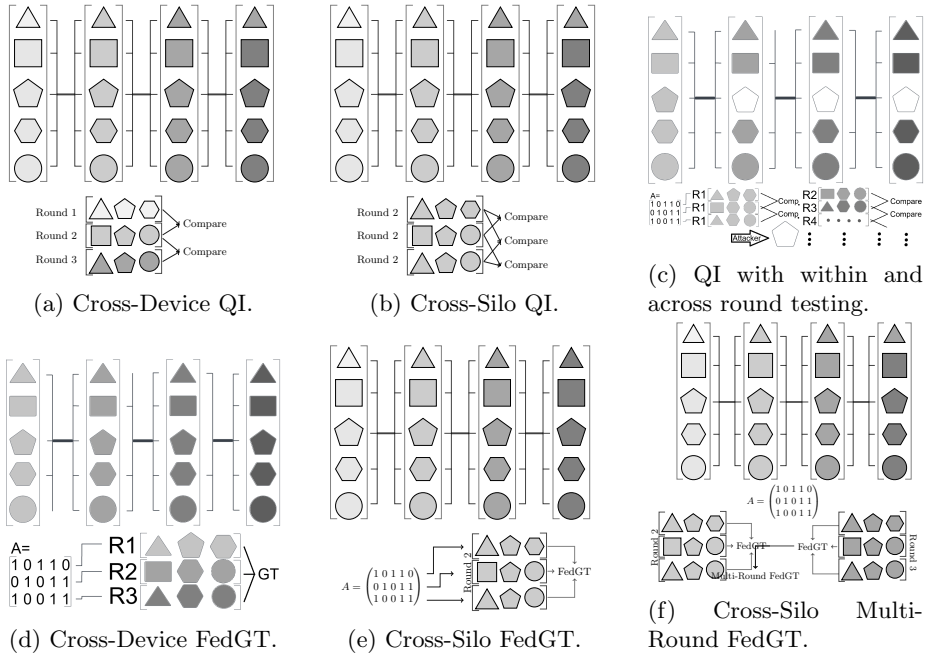

Identification of malicious clients in a privacy-preserving manner was addressed in~\cite{xhemrishi2023fedgt}, where the authors proposed FedGT.
Inspired by group testing, FedGT~\cite{xhemrishi2023fedgt} enables  MD under SA. The key idea is to group clients into overlapping groups, with each group performing SA. The server only receives the aggregated model updates from each group. By carefully designing the group structure and leveraging a decoding algorithm, the server can identify malicious clients based on the aggregated group outputs.  As illustrated in Fig.~\ref{fig:sil-gt}, FedGT was proposed for single-round testing in a cross-silo FL setting (in the illustrated example, only in round 2). Clients are grouped into overlapping groups according to an assignment matrix $\bA$.  After the group aggregates are collected, the server applies a testing algorithm followed by a decoding step to determine which clients are likely to be malicious.

The primary goal of FedGT is not to identify malicious clients, but to do so in order to achieve high good model utility even in the presence of malicious clients.  In extensive experiments, FedGT outperformed state-of-the-art private robust aggregation methods~\cite{pillutla2022robust,BREA}\textemdash which do not support the ability to identify malicious clients\textemdash in terms of global model utility and communication efficiency. As noted in~\cite{xhemrishi2023fedgt}, FedGT's MD performance could be further improved by extending it to  a multi-round testing scheme, which we pursue in this work. 

However, performing this strategy only once can make the impact of false alarms quite high, especially in a non-IID data distribution setting. Moreover, it might even risks that not all malicious clients are flagged as such, especially in a relatively large number of malicious clients regime.

\subsubsection*{Baseline}

As a baseline for MD, we use \textit{cosine similarity} (COS). COS is a versatile tool in machine learning, applied to both accelerate convergence~\cite{wu2021fast} and defend against adversarial attacks~\cite{zhu2024byzantine}. Here, we assign scores to clients based on the similarity between their local models and the global aggregated gradient. While this approach is simplistic, it is highly efficient, requiring only the computation of an inner (dot) product. However, in its standard form, COS is not privacy-preserving since it depends on client-specific models. Nevertheless, its computational simplicity allows it to be implemented using homomorphic encryption (HE)~\cite{acar2018survey}. 

\subsection{Privacy-Preserving Contribution Evaluation}

CE can be broadly categorized into three disciplines~\cite{rozemberczki2022shapley}: Explainability~\cite{gunning2019xai}, which assigns importance scores to individual data features, data evaluation~\cite{wang1996beyond}, which assesses the contribution of individual data points, and contribution scoring~\cite{wang2019measure}, which evaluates the impact of entire datasets corresponding to FL clients. Within this work, we will focus on the latter. 

Another distinctive aspect of CE is the reliance on an external test dataset. Some schemes, such as gradient similarity-based approaches~\cite{xu2021gradient}, avoid test sets and instead compute distance metrics between local and global models or gradients. The assumption is that clients whose updates closely align with the global model contribute more. In this paper, we assume the availability of an external test dataset. 

A flagship technique for CE is the Shapley value~\cite{winter2002shapley}, derived from cooperative game theory. It uniquely satisfies four key axioms, 
making it the only theoretically fair method for distributing rewards among players. The score of a client is computed by averaging its marginal contributions across all possible subsets of other clients. Yet, this computation is infeasible in real-world settings due to its exponential complexity. Consequently, several model-agnostic approximations have been developed but most rely on individual-level information (such as the gradients and model updates). This exposes client data to potential privacy risks, an aspect largely overlooked in existing literature. 

There are only a handful of work dealing with privacy-preserving CE. In~\cite{zheng2022secure}, the authors proposed a multi-server solution relying on encryption, while in ~\cite{watson2022differentially} they utilized DP, and ~\cite{ma2021transparent} built a solution on the blockchain technology. Regarding marginal contributions, in~\cite{pejoI1I,pejoEEE}, the authors proposed new techniques where the clients either conduct a self-evaluation or they evaluate everybody else, respectively. This work improves on~\cite{pejo2023quality} where the authors proposed \textit{Quality Inference} (QI), a group comparison solution that takes advantage of the client selection process in cross-device FL. 

\subsubsection*{Quality Inference}

QI operates within an honest-but-curious cross-device setting, where both the aggregator server and participants strictly adhere to the FL protocol. In an ideal world, during learning, it is expected that 1) the model improves every round, but 2) the rate of improvement decreases every round. QI captures deviations from these patterns by scoring the participating clients. In its core, as illustrated by Fig.~\ref{fig:dev-qi}, QI relies on three scoring rules: \emph{The good}, \emph{the bad}, and \emph{the ugly}. The authors of~\cite{pejo2023quality} experimented with various combinations of these rules as well as scaling the scores, but these had a limited effect. Here, we will consider the basic setting where equal weight is assigned to these unit scores. In more detail, the clients are scored according to the good and the bad 
 (if the current round improved the model more than the previous round, the selected clients for this and the previous round gain $+1$ and $-1$, respectively) and the ugly (if the current round does not improves the model, the selected clients gain $-1$) rules. 

 \subsubsection*{Baseline}

Within this work we will use the well-known Leave-One-Out (LOO) as our baseline. LOO is widely used in machine learning, including stability~\cite{evgeniou2004leave} and fairness~\cite{black2021leave}. LOO measures the marginal difference between the global model’s performance with and without a single client’s update. Hence, similarly to COS, LOO is also not inherently privacy-preserving, but its complexity is linear, which makes its computation feasible with HE. 
\section{\textsc{FedGT} and \textsc{QI} for MD and CE}
\label{sec:model}

In this section, we cross-validate FedGT and QI on MD and CE. The solutions for MD (aka detecting malicious clients) and CE (aka scoring participants) overlap and are somewhat interchangeable. While the original goal for FedGT was the former, QI aimed at the latter, so it is ambiguous in which domain they should be evaluated or measured. We envision several comparisons and combinations, and for brevity reasons, Fig.~\ref{fig:illustrate} illustrates a few of them. 

\begin{table}[!t]
    \centering
    \caption{Notations used in the paper for FedGT and QI.\vspace{2ex}}
    \begin{tabular}{cc !{\vrule width 1pt} p{9cm}}
        & Symbol & Description \\
        \Xhline{2\arrayrulewidth}
        \multirow{3}{*}{\rotatebox[origin=c]{90}{FL}} & $N$ \& T & Number of clients \& Number of epochs \\
        &$M^t_n$ & Client $n$'s locally trained model in round $t$ \\
        &$M^t$ & Aggregated model of clients in round $t$ \\
        \Xhline{1.5\arrayrulewidth}
        \multirow{4}{*}{\rotatebox[origin=c]{90}{Device}} &$S_{N\times T}$ & Binary client selection matrix for each round \\
        &$K$ & Number of clients participating in a round \\
        &$\mathcal{S}^t_n$ & Data Shapley of client $n$ based on the first $t$ round \\
        &$\mathcal{Q}^t_n$ & QI score of client $n$ based on the first $t$ rounds \\
        \Xhline{1.5\arrayrulewidth}
        \multirow{5}{*}{\rotatebox[origin=c]{90}{Silo}} &$\tau$ & Training round where the FedGT tests are performed \\
        &$L$ \& $k$ & Number of groups and their size for testing \\
        &$A_{N\times L}$ & Assignment matrix for in-round testing \\
        &$\hat{M}_l$ & Aggregated model of clients in test group $l$ \\
        &$\mathcal{G}_n^\tau$ & FedGT score of client $n$ based on round $\tau$ \\
       \Xhline{2\arrayrulewidth}
    \end{tabular}
    \label{tab:notations}
\end{table}

\subsubsection*{Notations}

In Table~\ref{tab:notations}, we summarize the basic notations for FedGT and QI, such as $M^t_n$ (model trained locally by client $n$ in round $t$), $M^t$ (aggregated model in round $t$ determined via selection matrix $S$ for cross-device FL), and $\hat{M}_l$ (aggregated model of group $l$ determined via assignment matrix $A$). $K$ and $k$ corresponds to the number of clients affiliated with $M^t$ and $\hat{M}_l$, and as usual, $N$ and $T$ stands for the overall number of clients and iterations, respectively. $\tau$ is the round in which the testing is utilized, and $L$ is the number of groups for the testing (e.g., for FedGT) while $\mathcal{S}_n$, $\mathcal{Q}_n$, and $\mathcal{G}_n$ are the scores of client $n$ based on Data Shapley, Quality Inference, and Group Testing, respectively. 

\subsection{Cross-Silo Federated Learning}

Focusing on the cross silo setting, we adopt QI and compare it with FedGT for both MD and CE tasks, and also propose a multi-round variant for both of them. 

\subsubsection{Adopting QI for Cross-Silo FL}

As illustrated in Fig.~\ref{fig:dev-qi}, QI performs comparisons between groups defined by the random selection matrix $S$, where each group has size $K$. However, in a cross-silo setting, all clients participate in every round, making the original QI approach inapplicable. FedGT resolves this by creating subgroups of clients based on the assignment matrix $A$ (as shown in Fig.~\ref{fig:sil-gt}), which is designed based on error-correcting codes. Building on this idea, QI can also be adapted to use these structured groups, and perform pairwise comparisons between them, as shown in Fig.~\ref{fig:sil-qi}. Rather than comparing randomly-formed groups across different rounds, comparisons are now made between carefully designed groups  within the same round. This adaptation shifts from performing $2\cdot (T-1)$ temporal comparisons to $L\cdot(L-1)$ spatial comparisons, leveraging the group structure for more effective evaluation in the cross-silo setting.

\subsubsection{Using FedGT for CE}

FedGT is originally designed to flag attackers using an optimal soft-decoding algorithm. This decoder utilizes likelihood ratios, which are soft labels for a binary outcome (malicious or not) and therefore, might be suitable for CE: the smaller or bigger the likelihood, presumably the higher or lower the corresponding client's data quality. 

\subsubsection{Multi-Round}

Compared to QI, FedGT relies on a single round. Extending it to multiple rounds (as illustrated in Fig.~\ref{fig:sil-multi}) would allow for more information regarding clients behavior in comparison with the single-round testing. Thus, it is expected that the MD performance would increase when the likelihoods are aggregated across test rounds. This extension implies that $\tau$ is a set (rather than a number) where the group testing commences. We envision three strategies for forming the assignment matrix across rounds: 
\begin{itemize}
\item \textbf{Same}: Use the same test groups across all rounds, i.e., $A^{\tau_i}=\bA^{\tau_j}$ for all $\tau_i, \tau_j\in\tau$.

\item \textbf{Prefixed}: Use different but predetermined test groups for each round,  where $A^{\tau_i}$ depends only on the round index $i$ for all $\tau_i\in\tau$. 

\item \textbf{Adaptive}: Use adaptive test groups for each round, where $A^{\tau_i}$ depends on $A^{\tau_j}$ and on $\mathcal{G}^{\tau_{j}}$ for all $\tau_j<\tau_i$. 
\end{itemize}

In this work, we focus on the Prefixed strategy, which we refer to as multi-round FedGT (MR-FedGT). Experimental results showed that the Same strategy provided no improvement, while exploration of the Adaptive strategy is left for future work.

Alternatively, QI can also be extended to the multi-round setting. In this case,  a single group defined by the assignment matrix $A^{\tau_i}$ is compared with the other $L-1$ groups from the same round (within-round comparison, as shown in Fig.~\ref{fig:sil-qi}) as well as with the $2\cdot L$ subgroups defined by $A^{\tau_{i-1}}$ and $A^{\tau_{i+1}}$ (across-round comparison, as illustrated in Fig.~\ref{fig:dev-qi}).  Since this scheme considers a larger set of group comparisons\textemdash unlike FedGT, which operates only  within rounds\textemdash  it is expected to achieve superior performance. We refer to this extension as multi-round QI (MR-QI). Similarly to MR-FedGT, MR-QI adopts the Prefixed strategy for constructing multiple assignment matrices across rounds. 

The assignment matrix $A$ is designed to ensure that no client’s raw model $M_n^t\,, \forall n \in [N]$ can be inferred from any linear combination of the tested groups within a single round. However, as the learning process progresses toward convergence, models from consecutive rounds become increasingly similar, i.e., models $M_n^t$ and $M_n^{t+1}$ are nearly identical when $t$ is large. Therefore, when considering both $A^{\tau_i}$ and $A^{\tau_{i+1}}$ simultaneously, if $\tau_i$ is sufficiently large and $\tau_{i+1}$ is  close to it, the privacy guarantees may deteriorate. Designing assignment matrices that account for such subtle correlations\textemdash whether under Prefixed or Adaptive strategies\textemdash remains an open problem and is left for future work.


\subsection{Cross-Device Federated Learning}

Focusing on the cross device setting, we adopt FedGT and compare it with QI for both MD and CE tasks and also propose a comprehensive solution (to tackle both MD and CE) based on their combinations.

\subsubsection*{Adopting FedGT to Cross Device}

As illustrated on Fig.~\ref{fig:sil-gt}, FedGT utilizes the groups defined by the assignment matrix $A$, where the size of each set is $k$. In contrast, in a cross-device setting, $K$ clients participate in each round; hence, the vanilla FedGT is not applicable. QI resolves this by comparing the random groups created by the client selection mechanism (as shown in Fig.~\ref{fig:dev-qi}). Hence, FedGT can also use the groups defined by the round: the clients are grouped as per the rows of the assignment matrix in different rounds and only later are the groups tested, after the last sampling round has been collected. This strategy requires that the client sampling is performed not randomly, but in a structured manner following $A$, which may introduce bias in the model. By chance, the random selection process might define the groups as $A$, but they could correspond to very different iterations; thus, comparing them carries little meaning. Thus, a trade-off between the model's performance and the accuracy of MD and CE exists. 

\subsection{Comprehensive Solution}

Taken the best of both worlds, to provide a comprehensive solution, a sub-group-based technique within an early round (see Fig.~\ref{fig:sil-qi} and~\ref{fig:sil-gt}) can be used for MD and and a group based technique across the following round (see Fig.~\ref{fig:dev-qi} and~\ref{fig:dev-gt}) for CE sequentially, as shown in Fig.~\ref{fig:dev-comb}. We originally anticipated that FedGT will be used for MD and QI for CE, but our empirical findings revealed that latter is superior for both tasks, so the illustration encapsulates that. Note that if there is no communication between clients (which is reasonable in cross-device setting), the testing is stealthy, i.e., the corresponding round number is hidden. Thus, the attackers cannot change their strategy temporarily. 

Similarly to~\cite{xhemrishi2023fedgt}, the MD detection should be done in the first few rounds except the first, when the model improves the most, so it is hard to differentiate the clients. Alternatively, when a pretrained model is used, the first round can also be considered for testing. In that round all $N$ clients should participate (instead of the $K$), so the subgroup based detection technique can be used. This does neither introduce any bias in the training nor does compromise privacy if $k\geq K$, which can be arranged easily. Finally, the suspicious clients should be discarded from training. Afterwards the training proceeds with any client selection method (only considering the presumably benign clients), and CE scores are assigned based on the adjacent round's performance improvements.
\section{Experiments}
\label{sec:exp}

This section details our experimental setup and key findings, i.e., we verify our adaptation of QI for the cross silo setting, experiment with FedGT for CE, and measure the performance improvement of multi-round techniques. 

\subsubsection*{Setup}

We conduct experiments for a classification problem over image datasets, namely the well-known CIFAR-10
and ISIC2019 datasets. The latter contains skin lesions and is tailored for cross-silo FL~\cite{Flamby,xhemrishi2023fedgt}. We train Resnet-18 (with $0.05$ as learning rate, $0.9$ as momentum, and $0.001$ as weight decay) and EfficientNet-B0 (pretrained on ImageNet dataset, with $0.0005$ as learning rate, $0.9$ as momentum, and $0.0001$ as weight decay) for CIFAR-10 and ISIC2019, respectively. We simulate FL using 15 clients; ISIC2019 corresponds to non-IID scenario (as it is inherited real-world heterogeneity) while we utilize two distribution scenarios for CIFAR-10: IID and non-IID using Dirichlet distribution with a parameter of $0.5$. Measuring ground-truth contribution scores is difficult (as the Shapley value is computationally infeasible), so following~\cite{pejo2023quality}, we introduce noise linearly to the labels to create variation: for client $n$ each label is changed with probability $\frac{n}{N+1}$. To ensure statistically significant results, we fix the data splits and the injected noise and repeat each training process ten times. 

The experiments consist of 20 federated rounds where the clients perform $5$ (for CIFAR-10) or $ 1$ local epochs (for ISIC) using stochastic gradient descent optimization with $128$ (for CIFAR-10) and $ 64$ (for ISIC) as batch sizes. For experiments over CIFAR-10, we use the cross entropy loss, while for ISIC, we use the focal loss since it is shown to perform better over unbalanced datasets. Regarding the attack scenarios, we employ an untargeted label-flipping data-poisoning attack, offsetting the labels of the training data by shifting them by one. The components (testing algorithm, validation dataset size, decoder etc) of FedGT and the hyper-parameters are taken from the original FedGT paper~\cite{xhemrishi2023fedgt}.

As a baseline for MD, we use COS. While more advanced MD strategies exist, most are incompatible with privacy-preserving settings. Based on the obtained detection scores, we utilized a clustering approach to separate the suspected attackers and the anticipated benign clients based on agglomerative Clustering technique. 

For CE, we use LOO as our baseline. Note, that the scores are expected to be on different scales: FedGT scores are unbounded (as they rely on likelihood ratios), QI is mostly negative (as the majority of scoring rules are punishing the participants), and for LOO some scores are above while some are below zero (similarly to the Shapley Value). For fair comparison, we transform them via minimum offset correction (shifts all scores by subtracting the minimum score from each) and normalization (the scores are divided by the sum).

Our implementation relied on PyTorch and it is important to note that we do not employ cryptographic techniques in our implementation; rather, we use these simple computations (as baseline) that ensure their practical applicability with HE. 

\subsection{Cross Silo FL}

In this subsection we verify our adaptation of QI for the cross silo setting, experiment with FedGT for CE, and measure the performance improvement of multi-round techniques. 

\subsubsection{Adopting QI for Cross Silo}

\begin{table}[!t]
    \centering
    \caption{Detection performance (F1-score) of QI and FedGT where 1, 3, 5, and 7 clients are attackers (out of 15). The testing is performed in every round and the measurement per round is aggregated. The tabulated results are measured at the last training round. 1R denotes the single round approach (the maximum throughout all the rounds), while MR refers to multi-round approach.\vspace{2ex}}
    \begin{tabular}{ccc !{\vrule width 1pt} cc|c !{\vrule width 1pt} cc}
        \multicolumn{3}{c !{\vrule width 1pt} }{Cross-Silo} & MR-QI & MR-FedGT & COS & 1R-QI & 1R-FedGT \\
        \Xhline{2\arrayrulewidth}
        \multirow{8}{*}{\rotatebox{90}{CIFAR-10}} & \multirow{4}{*}{\rotatebox{90}{IID}} & 1 & $1.00\pm0.00$ & $1.00\pm0.00$ & $1.00 \pm 0.00$ & $0.70 \pm 0.46$ & $1.00 \pm 0.00$ \\
        &  & 3 & $0.97 \pm 0.06$ & $1.00\pm0.00$ & $1.00 \pm 0.00$ & $0.67 \pm 0.26$ & $0.77\pm0.21$ \\
        &  &  5 & $0.98 \pm 0.04$ & $0.96 \pm 0.05$ & $1.00 \pm 0.00$ & $0.74 \pm 0.13$ & $0.58\pm0.14$ \\
        &  &  7 & $0.98 \pm 0.05$ & $0.98 \pm 0.05$ & $1.00 \pm 0.00$ & $0.79 \pm 0.07$ & $0.74\pm0.09$ \\
        \cline{2-8}
        & \multirow{4}{*}{\rotatebox{90}{non-IID}} & 1 & \best{$0.72 \pm 0.36$} & $0.42 \pm 0.35$ & $0.61 \pm 0.34$ & $0.50 \pm 0.50$ & $0.30\pm0.46$  \\
        &  & 3 & \best{$0.87 \pm 0.19$} & $0.54 \pm 0.21$ & $0.74 \pm 0.18$ &  $0.67 \pm 0.26$ & $0.37\pm0.28$ \\
        &  & 5 & \best{$0.84 \pm 0.22$} & $0.50 \pm 0.19$ & $0.57 \pm 0.25$ & $0.70 \pm 0.18$ & $0.44\pm0.17$ \\
        &  & 7 & \best{$0.88 \pm 0.17$} & $0.70 \pm 0.20$ & $0.38 \pm 0.29$ & $0.73 \pm 0.08$ & $0.59\pm0.13$ \\
        \Xhline{1.5\arrayrulewidth}
        \multirow{4}{*}{\rotatebox{90}{ISIC19}} & \multirow{4}{*}{\rotatebox{90}{non-IID}} & 1 & $0.88 \pm 0.26$ & $0.90\pm0.30$ & $1.00 \pm 0.00$ & $0.60 \pm 0.49$ & $0.40\pm0.49$\\
        &  & 3 & \best{$0.80 \pm 0.17$} & $0.60 \pm 0.19$ & $0.96 \pm 0.08$ & $0.57 \pm 0.21$ & $0.27\pm0.25$ \\
        &  & 5 & \best{$0.91 \pm 0.07$} & $0.61 \pm 0.19$ & $0.89 \pm 0.05$ & $0.66 \pm 0.13$ & $0.44\pm0.25$ \\
        &  & 7 & \best{$0.91 \pm 0.06$} & $0.59 \pm 0.11$ & $0.68 \pm 0.08$ &  $0.76 \pm 0.07$ & $0.56\pm0.23$ \\
        \Xhline{2\arrayrulewidth}
    \end{tabular}
    \label{tab:CS-MD-MR}
\end{table}

While the paper proposing QI~\cite{pejo2023quality} did consider MD, it only did so in the cross-device setting for IID clients, while FedGT~\cite{xhemrishi2023fedgt}, only considered the cross-silo setting. 
This paper investigates both the IID and non-IID settings and compares the schemes with COS. Our results are summarized on the right side of Table~\ref{tab:CS-MD-MR} for various numbers of attacks. These values suggest a correlation between the number of attackers and the detection performance, as more attackers correspond to better detection (except for FedGT with IID). Comparing the last two columns, we can conclude that the simplistic QI rules outperform the more involved FedGT scheme when the data distribution is non-IID. At the same time, the comparison regarding the IID setting is inconclusive. 

\subsubsection{Multi-round MD}

 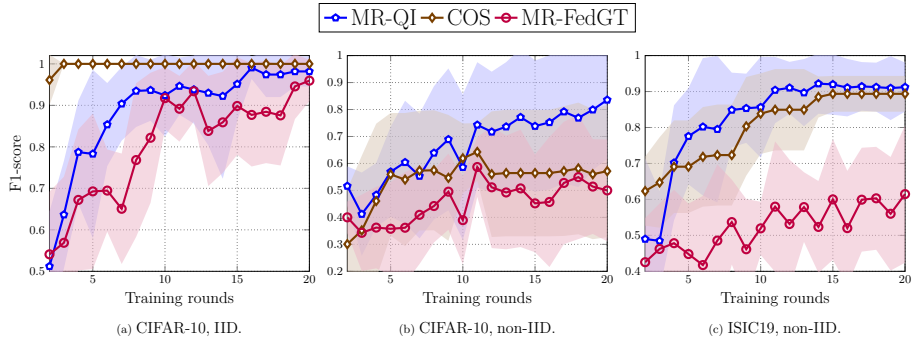
\begin{figure*}[!t]
     \centering
     \resizebox{\textwidth}{!}{\definecolor{chocolate}{rgb}{0.48, 0.25, 0.0}
\definecolor{amber}{rgb}{1.0, 0.75, 0.0}
\begin{tikzpicture}

\begin{groupplot}[ 
    group style={
        group size=3 by 1,
        horizontal sep=1cm
   },
]
\nextgroupplot
[
    grid = both, 
    grid style={dotted,draw=black!90},
    tick label style={/pgf/number format/fixed},
    xmode = linear, 
    ymode = linear, 
    ymax = 1.02, 
    ymin = 0.5, 
    xmax = 20, 
    xmin = 2, 
    xlabel = \large Training rounds, 
    ylabel = \large F1-score,
    ylabel near ticks,
    xlabel near ticks
]
    \addplot[blue, ultra thick, solid, mark=pentagon*, mark options={fill=white} ,mark size=2.5pt] table [x index = {0}, y index={3}, col sep=comma]
    {CIFAR10_homo_m_5.tex};
    
    \addplot[chocolate, ultra thick, solid, mark=diamond*, mark options = {fill = white}, mark size=3pt] table [x index = {0}, y index={1}, col sep=comma]{CIFAR10_homo_m_5.tex};
    
    \addplot[purple, ultra thick, solid, mark=o, mark options = {fill = white}, mark size=3pt] table [x index = {0}, y index={2}, col sep=comma]{CIFAR10_homo_m_5.tex};

        \addplot+[name path=QI-low, draw=none, mark=none] 
    table [x index = {0}, y index={8}, col sep=comma]
    {CIFAR10_homo_m_5.tex};
    \addplot+[name path=QI-high, draw=none, mark=none] 
    table [x index = {0}, y index={9}, col sep=comma]
    {CIFAR10_homo_m_5.tex};
    \addplot [
        blue!30, 
        opacity=0.5
    ] fill between[
        of=QI-low and QI-high,
    ];

    \addplot+[name path=COS-low, draw=none, mark=none] 
    table [x index = {0}, y index={4}, col sep=comma]
    {CIFAR10_homo_m_5.tex};
    \addplot+[name path=COS-high, draw=none, mark=none] 
    table [x index = {0}, y index={5}, col sep=comma]
    {CIFAR10_homo_m_5.tex};
    \addplot [
        chocolate!30, 
        opacity=0.5
    ] fill between[
        of=COS-low and COS-high,
    ];

    \addplot+[name path=GT-low, draw=none, mark=none] 
    table [x index = {0}, y index={6}, col sep=comma]
    {CIFAR10_homo_m_5.tex};
    \addplot+[name path=GT-high, draw=none, mark=none] 
    table [x index = {0}, y index={7}, col sep=comma]
    {CIFAR10_homo_m_5.tex};
    \addplot [
        purple!30, 
        opacity=0.5
    ] fill between[
        of=GT-low and GT-high,
    ];
    \nextgroupplot
    [
        grid = both, 
        grid style ={dotted,draw=black!90},
        tick label style ={/pgf/number format/fixed},
        xmode = linear, 
        ymode = linear, 
        ymax = 1.0, 
        ymin = 0.2, 
        xmax = 20, 
        xmin = 2, 
        xlabel = \large Training rounds, 
        legend style =
        {
            legend columns=6,
            fill=white,
            draw=black,
            anchor=south,
            legend cell align=left,
            align=left
        },
        legend to name=acc_legend_round_acc,
        ytick={0.2, 0.3, 0.4, 0.5, 0.6, 0.7, 0.8, 0.9, 1},
        xlabel near ticks
    ]

    \addplot[blue, ultra thick, solid, mark=pentagon*, mark options={fill=white} ,mark size=2.5pt] table [x index = {0}, y index={3}, col sep=comma]
    {CIFAR10_hetero_m_5.tex};
    \addlegendentry{\LARGE MR-QI}

    \addplot[chocolate, ultra thick, solid, mark=diamond*, mark options = {fill = white}, mark size=3pt] table [x index = {0}, y index={1}, col sep=comma]{CIFAR10_hetero_m_5.tex};
    \addlegendentry{\LARGE COS}

    \addplot[purple, ultra thick, solid, mark=o, mark options = {fill = white}, mark size=3pt] table [x index = {0}, y index={2}, col sep=comma]{CIFAR10_hetero_m_5.tex};
    \addlegendentry{\LARGE MR-FedGT}

    \addplot+[name path=QI-low, draw=none, mark=none] 
    table [x index = {0}, y index={8}, col sep=comma]
    {CIFAR10_hetero_m_5.tex};
    \addplot+[name path=QI-high, draw=none, mark=none] 
    table [x index = {0}, y index={9}, col sep=comma]
    {CIFAR10_hetero_m_5.tex};
    \addplot [
        blue!30, 
        opacity=0.5
    ] fill between[
        of=QI-low and QI-high,
    ];

    \addplot+[name path=COS-low, draw=none, mark=none] 
    table [x index = {0}, y index={4}, col sep=comma]
    {CIFAR10_hetero_m_5.tex};
    \addplot+[name path=COS-high, draw=none, mark=none] 
    table [x index = {0}, y index={5}, col sep=comma]
    {CIFAR10_hetero_m_5.tex};
    \addplot [
        chocolate!30, 
        opacity=0.5
    ] fill between[
        of=COS-low and COS-high,
    ];

    \addplot+[name path=GT-low, draw=none, mark=none] 
    table [x index = {0}, y index={6}, col sep=comma]
    {CIFAR10_hetero_m_5.tex};
    \addplot+[name path=GT-high, draw=none, mark=none] 
    table [x index = {0}, y index={7}, col sep=comma]
    {CIFAR10_hetero_m_5.tex};
    \addplot [
        purple!30, 
        opacity=0.5
    ] fill between[
        of=GT-low and GT-high,
    ];
    

    

    \nextgroupplot
    [
        grid = both, 
        grid style={dotted,draw=black!90},
        tick label style={/pgf/number format/fixed},
        xmode=linear, 
        ymode=linear, 
        ymax=1.0, 
        ymin = 0.4, 
        xmax = 20, 
        xmin=2, 
        xlabel= \large Training rounds, 
        ytick={0.2, 0.3, 0.4, 0.5, 0.6, 0.7, 0.8, 0.9, 1},
        xlabel near ticks
    ]

    \addplot[blue, ultra thick, solid, mark=pentagon*, mark options={fill=white} ,mark size=2.5pt] table [x index = {0}, y index={3}, col sep=comma]
    {ISIC_homo_m_5.tex};
    
    \addplot[chocolate, ultra thick, solid, mark=diamond*, mark options = {fill = white}, mark size=3pt] table [x index = {0}, y index={1}, col sep=comma]{ISIC_homo_m_5.tex};
    
    \addplot[purple, ultra thick, solid, mark=o, mark options = {fill = white}, mark size=3pt] table [x index = {0}, y index={2}, col sep=comma]{ISIC_homo_m_5.tex};

    \addplot+[name path=QI-low, draw=none, mark=none] 
    table [x index = {0}, y index={8}, col sep=comma]
    {ISIC_homo_m_5.tex};
    \addplot+[name path=QI-high, draw=none, mark=none] 
    table [x index = {0}, y index={9}, col sep=comma]
    {ISIC_homo_m_5.tex};
    \addplot [
        blue!30, 
        opacity=0.5
    ] fill between[
        of=QI-low and QI-high,
    ];

    \addplot+[name path=COS-low, draw=none, mark=none] 
    table [x index = {0}, y index={4}, col sep=comma]
    {ISIC_homo_m_5.tex};
    \addplot+[name path=COS-high, draw=none, mark=none] 
    table [x index = {0}, y index={5}, col sep=comma]
    {ISIC_homo_m_5.tex};
    \addplot [
        chocolate!30, 
        opacity=0.5
    ] fill between[
        of=COS-low and COS-high,
    ];

    \addplot+[name path=GT-low, draw=none, mark=none] 
    table [x index = {0}, y index={6}, col sep=comma]
    {ISIC_homo_m_5.tex};
    \addplot+[name path=GT-high, draw=none, mark=none] 
    table [x index = {0}, y index={7}, col sep=comma]
    {ISIC_homo_m_5.tex};
    \addplot [
        purple!30, 
        opacity=0.5
    ] fill between[
        of=GT-low and GT-high,
    ];

\end{groupplot}
\node[above] at ([yshift=5mm]group c2r1.north)
{\pgfplotslegendfromname{acc_legend_round_acc}};

\node[text width=6cm,align=center,anchor=north] at ([yshift=-10mm]group c1r1.south) {\subcaption{\large CIFAR-10, IID.}\label{fig:CIFAR_iid}};
\node[text width=6cm,align=center,anchor=north] at ([yshift=-10mm]group c2r1.south) {\subcaption{\large CIFAR-10, non-IID.}\label{fig:CIFAR_non_iid}};
\node[text width=6cm,align=center,anchor=north] at ([yshift=-10mm]group c3r1.south) {\subcaption{\large ISIC19, non-IID.}\label{fig:ISIC}};

\end{tikzpicture}%
}
     \caption{Detection performance (F1-score) of QI, FedGT and COS where $5$ clients are malicious (out of $15$) versus the communication rounds. The highlighted area represents the standard deviation obtained in our experiments.}
      \label{fig:multi}
 \end{figure*}

In multi-round MD, the testing is performed at each training round (except the first). 
We use QI with both across- and within-round comparisons, while for FedGT, we restrict ourselves to in-round comparisons as input for the testing algorithm. We tried to adopt FedGT for the Cross-Device setting (to enable across round testing too) by selecting one prefixed groups for each training round, but even with a Taylor-based interpolation (to normalize the improvement from different rounds to the same expected scale) FedGT is failing to provide meaningful results, so we discarded this option. 

We present our results in the left side of Table~\ref{tab:CS-MD-MR}. It is clear that performing the tests in multiple round significantly enhances both schemes, as the values on the left (multi-round) are consistently larger than on the right (single-round). However, the trend (more attackers imply better detection) visible for 1R is not holding for MR. Regarding the baseline COS mechanism (which we also applied in multiple round by accumulating the computed scores), it does slightly outperform the two privacy-preserving schemes in the IID setting. On the other hand, the more diverse the client's data distributions (ISIC can be considered low imbalance, CIFAR with Dirichlet(0.5) can be considered high imbalance), the worst COS performs. In contrast, QI seems to be robust against such changes and clearly outperforms both COS and FedGT in that setting (middle lines of the table). These results suggests, that CE schemes (such as QI) could be appropriate for MD as well; the fine-grained scoring can be turned into [0,1] (as benign and malicious) accurately.

Fig.~\ref{fig:multi} showcases how the detection accuracy behaves in respect to training rounds. Similar to the results in Table~\ref{tab:CS-MD-MR} the non-private scheme COS achieves almost perfect detection and it achieves it very fast (third round, see Figure~\ref{fig:CIFAR_iid}). However, the proposed multi-round QI performs well and reaches F1-score at seventh round. For experiments over CIFAR-10, where the client data is distributed according to Dirichlet with parameter $0.5$ (Figure~\ref{fig:CIFAR_non_iid}), the proposed multi-round QI outperforms COS in almost every round. At the eleventh round, QI reaches a F1-score above $0.7$. Due to the heterogeneity, all schemes suffer from a relatively high standard deviation. A similar pattern is observed for experiments over ISIC19 datasets, as one can see from Figure~\ref{fig:ISIC}. Except for the first two measures (recall that the measurement starts at the second round), the multi-round QI outperforms all the other schemes. However, it is important to note that the proposed multi-round FedGT improved compared to its original single-round form, but is outperformed by both COS and the proposed multi-round QI.

\subsubsection{Using MR-FedGT for CE}

\begin{table}[!t]
    \centering
    \caption{Scoring performance (the distance of the score vectors $L_2$ and their ordering differences $\phi$) of QI, FedGT, and LOO. The testing is performed in every round and the tabulated results are measured at the last training round.\vspace{2ex}}
    \begin{tabular}{ccc !{\vrule width 1pt} ccc}
         \multicolumn{3}{c!{\vrule width 1pt}}{Cross-Silo} & MR-QI & MR-FedGT & LOO \\
        \Xhline{2\arrayrulewidth}
        \multirow{4}{*}{\rotatebox{90}{CIFAR}}&\multirow{2}{*}{IID} & $L_2$ & $0.015\pm 0.002$ & $0.029\pm 0.007$ & $0.036\pm 0.012$ \\
        && $\phi$ & $0.944\pm0.027$ & $0.761\pm0.094$ &  $0.996\pm0.004$ \\
        \cline{2-6}
        &\multirow{2}{*}{non-IID} & $L_2$ & $0.025\pm 0.009$ & $0.041\pm 0.007$ & $0.034\pm 0.007$ \\
        && $\phi$ & $0.791\pm0.146$ & $0.429\pm0.158$ & $0.606\pm0.246$ \\
        \Xhline{1.5\arrayrulewidth}
        \multirow{2}{*}{\rotatebox{90}{ISIC}}&\multirow{2}{*}{non-IID} & $L_2$ & $0.036\pm 0.007$ & $0.045\pm 0.009$ & $0.037\pm 0.005$ \\
        && $\phi$ & $0.675\pm0.097$ & $0.299\pm0.299$ & $0.582\pm0.111$ \\
        \Xhline{2\arrayrulewidth}
    \end{tabular}
    \label{tab:CS-CE}
\end{table}

The paper proposing FedGT~\cite{xhemrishi2023fedgt} only considered MD, so here we are testing the hypothesis whether an MD mechanism is appropriate for CE. This is less studied in the literature, while the opposite direction is prevalent, see for instance~\cite{sun2023shapleyfl}. Note we are injecting noise to the client labels as described earlier. Our results are summarized in Table~\ref{tab:CS-CE} where the privacy-preserving CE methods are compare to the utilized noise ratios, that are also transformed similarly to the other scores described previously. The table presents the $L_2$ errors where smaller is the better and the Spearman correlation coefficient $\phi$ where larger is the better. The first shows the absolute difference between the transformed scores, while the latter captures the ordering preservation of the clients: the metric ranges from $[-1,1]$, where positive values indicate strong agreement and values near zero imply little to no correlation. This is a standard technique in CE~\cite{jiang2023opendataval} to determine the accuracy of inferring the top and bottom performing clients. 
These results suggest that mechanisms tailored for MD (such as FedGT) are not necessarily applicable to CE as FedGT performs poorly in this task. It is indeed tricky to turn a classification (attacker vs non-attacker) into regression (client-scoring). On the other hand, QI does outperforms LOO both in the score ordering and in the score distances, showing that simple privacy-preserving solutions could outperform naive scoring techniques which rely on individual differences (hence, need to be combined with encryption for any privacy guarantee).

\section{Conclusion}
\label{sec:conc}

We improved and combined two existing schemes, QI and FedGT, into a scheme that we coin multi-round QI. The proposed scheme outperforms their previous versions and baselines for both misbehavior detection and contribution evaluation for a cross-silo FL scenario. The proposed scheme achieves very good performance, especially when the data is non-IID, while still preserving in-round clients privacy, due to compatibility with secure aggregation.

\textit{Limitation \& Future Work:} Our work has several limitations. Our empirical results are somewhat limited, as they assume only computer vision with a specific number of clients. In the future we plan to extend this work and also incorporate adaptive group testing, which is a particularly interesting direction. Moreover, we are currently investigating how to incorporate inter-round privacy protection for the clients, an attribute that was not the scope of this paper. 

\section*{Acknowledgments}
The authors are grateful to Gergely Biczók and Johan Östman for their comments on this work. 
This work was partially supported by the German Research Foundation (DFG) under Grant Agreement No. WA 3907/7-1, by the Swedish Research Council under grants 2020-03687
and 2023-05065, and by the Wallenberg AI, Autonomous Systems and Software Program (WASP) funded
by the Knut and Alice Wallenberg Foundation. 
Balázs Pejó was supported 1) by Project no. 145832 (implemented by the Ministry of Innovation and Technology from the NRDI Fund), 2) by the European Union project RRF-2.3.1-21-2022-00004 within the framework of the Artificial Intelligence National Laboratory, and 3) by the European Union (Grant Agreement Nr. 10109571, SECURED Project).

\bibliographystyle{ieeetr}
\bibliography{ref}

\end{document}